\documentclass[twocolumn,showpacs,floats,floatfix,superscriptaddress,aps,pra]{revtex4}
\usepackage{eurosym}
\usepackage{amsfonts}
\usepackage{amssymb}
\usepackage{amsmath}
\usepackage{calc}
\usepackage{graphicx}
\usepackage{bm}

\begin{document}

\author{Peter Chernev}
\affiliation{Center for Quantum Technologies, Department of Physics, Sofia University,
James Bourchier 5 blvd., 1164 Sofia, Bulgaria}
\author{Mouhamad Al-Mahmoud}
\affiliation{Center for Quantum Technologies, Department of Physics, Sofia University,
James Bourchier 5 blvd., 1164 Sofia, Bulgaria}
\author{Andon A. Rangelov}
\affiliation{Center for Quantum Technologies, Department of Physics, Sofia University,
James Bourchier 5 blvd., 1164 Sofia, Bulgaria}
\title{Universal composite phase gates with tunable target phase}
\date{\today }

\begin{abstract}
We present a systematic method for constructing \emph{universal} composite phase gates with a
continuously tunable target phase. Using a general Cayley--Klein parametrization of the
single-pulse propagator, we design gates from an \emph{even} number of nominal $\pi$ pulses and
derive analytic phase families by canceling, order by order in a small deviation parameter, the
leading contributions to the undesired off-diagonal element of the composite propagator,
independently of the dynamical phase. The resulting sequences provide intrinsic robustness against generic control imperfections and parameter fluctuations and remain valid for arbitrary pulse shapes. Numerical simulations in a standard
two-level model confirm high-order error suppression and demonstrate broad, flat high-fidelity
plateaus over wide ranges of simultaneous pulse-area and detuning errors, highlighting the
efficiency of the proposed \emph{universal} composite phase gates for resilient phase control in quantum
information processing.
\end{abstract}

\maketitle


\section{Introduction}

High-fidelity single- and two-qubit gates are essential for scalable quantum
information processing, where coherent control must be maintained over many
operations in the presence of imperfect experimental parameters~\cite{QI}.
Among the elementary gates, phase gates play a central role: they appear as
building blocks in a wide range of algorithms and protocols (including the
quantum Fourier transform), and they are repeatedly applied in phase-sensitive
subroutines such as Shor's factorization~\cite{Shor1,Shor2} and Grover's
search~\cite{Grover}. Consequently, the accuracy and robustness of phase
gates are critical for fault-tolerant quantum computation.

A phase gate changes the relative phase between two basis states without
changing their populations. For a fixed phase shift of $\pi$, a common
implementation uses a resonant $2\pi$ pulse that couples one qubit state to
an auxiliary level. Implementing an arbitrary target phase, however, requires
precise control of the interaction---often through the detuning, pulse area,
and timing---and is therefore susceptible to systematic errors. These issues
are compounded in realistic settings by additional perturbations such as
ac Stark shifts, spatial inhomogeneities, and unwanted frequency chirp.

Several physical mechanisms have been used to realize phase gates. Dynamic
phase gates~\cite{Cirac} can be implemented with a single off-resonant pulse
and are conceptually simple, but their performance depends sensitively on the
experimental parameters. Geometric phase gates~\cite{geometric1,geometric2,geometric3,geometric4}
may offer improved tolerance to certain fluctuations, at the cost of more
demanding control. Other approaches combine adiabatic passage with controlled
laser phases to stabilize the acquired phase~\cite{laser phases1,laser phases2}.
In all cases, the overarching goal is the same: to achieve phase control that
is simultaneously accurate, tunable, and robust against unknown variations in
the driving field.

Composite pulses (CPs) offer a simple and powerful route to robust control by
replacing a single pulse with a sequence of pulses with controlled relative
phases~\cite{Levitt79,Freeman80,Levitt86,Freeman97}. The key idea is that the
errors introduced by individual pulses interfere destructively, canceling the
leading sensitivity to imperfections while retaining high fidelity for the
desired operation. Conventional CP designs are typically tailored to correct
specific error types (e.g., pulse-area or detuning errors) and often assume
particular pulse shapes.

A major step beyond this paradigm is provided by \emph{universal} composite
pulses, introduced for robust population inversion by Genov \emph{et al.}~\cite{GenovUCP,GenovUCP2}.
Universal CPs are designed to compensate deviations in \emph{any} parameter of
the driving field (pulse amplitude and duration, detuning, Stark shifts,
unwanted chirp, etc.) and, importantly, they remain applicable for arbitrary
pulse envelopes. The universality is achieved by making no assumptions about
how these experimental parameters map onto the propagator of a single pulse.
Instead, one uses a general Cayley--Klein (St\"uckelberg) parametrization of
the single-pulse propagator and expands a suitable matrix element of the
composite propagator in a power series of a small parameter $\epsilon$ that quantifies
the deviation from the ideal transition. By choosing the CP phases so that the
coefficients of the lowest-order terms vanish for any dynamic phase,
one obtains sequences whose infidelity scales as a high power of $\epsilon$, thereby
producing a broad, flat high-fidelity plateau~\cite{GenovUCP,GenovUCP2}.

In the present work we bring this universal-CP philosophy to phase control.
Rather than constructing a phase gate from two identical composite $\pi$
pulses with a fixed relative shift---as in the composite phase-gate approach
of Torosov \emph{et al.}~\cite{Torosov} and its later developments~\cite{Gevorgyan}---we apply
the universality conditions \emph{directly} to the design of phase gates.
Specifically, we employ sequences of \emph{even $\pi$ pulses}, which
act as an elementary phase gate in the ideal limit, and we introduce and
optimize additional independent relative phases between each elementary
block. This provides sufficient freedom to (i) realize a continuously tunable
target phase and (ii) enforce universal cancellation of the leading error terms
in the composite-gate propagator. Crucially, this direct design yields shorter
sequences with equivalent robustness: our four-pulse gate achieves the same
high-order error suppression as prior six-pulse protocols. The resulting
\emph{universal composite phase gates} retain the key advantages of universal
CPs: robustness to arbitrary systematic field errors and applicability with any
pulse shape, while enabling high-fidelity tunable phase control suitable for
quantum information processing.


\section{Composite pulses}
\label{CP}

Consider a coherently driven two-state quantum system (qubit) in the general state
\begin{equation}
|\Psi(t)\rangle=c_{1}(t)\,|\psi_{1}\rangle+c_{2}(t)\,|\psi_{2}\rangle ,
\label{wave function}
\end{equation}
whose evolution is governed by the Schr\"odinger equation
\begin{equation}
i\hbar\,\partial_{t}\mathbf{c}(t)=\mathbf{H}(t)\,\mathbf{c}(t),
\end{equation}
with $\mathbf{c}(t)=[c_{1}(t),c_{2}(t)]^{T}$. Within the rotating-wave approximation
(RWA)~\cite{Allen-Eberly,Shore}, the Hamiltonian can be written as
\begin{equation}
\mathbf{H}(t)=\frac{\hbar}{2}
\begin{bmatrix}
-\Delta(t) & \Omega(t) \\
\Omega(t)  & \Delta(t)
\end{bmatrix},
\label{Hamiltonian}
\end{equation}
where $\Delta(t)=\omega_{0}-\omega(t)$ is the detuning between the driving-field
frequency and the Bohr transition frequency. The (generally time-dependent) Rabi
frequency $\Omega(t)$ characterizes the coupling strength; for an electric-dipole
transition $\Omega(t)=-\mathbf{d}\!\cdot\!\mathbf{E}(t)/\hbar$.

For a single pulse, the evolution from the initial time $t_i$ to the final time $t_f$
is described by a unitary propagator $\mathbf{U}$ such that $\mathbf{c}(t_f)=\mathbf{U}\,\mathbf{c}(t_i)$.
Following the universal composite-pulse formulation~\cite{GenovUCP,GenovUCP2}, we parametrize the
single-pulse propagator as
\begin{equation}
\mathbf{U}(\epsilon,\alpha,\beta)=
\begin{bmatrix}
\epsilon\,e^{i\alpha} & \sqrt{1-\epsilon^{2}}\,e^{i\beta} \\[4pt]
-\sqrt{1-\epsilon^{2}}\,e^{-i\beta} & \epsilon\,e^{-i\alpha}
\end{bmatrix},
\label{Uab}
\end{equation}
where $\epsilon$ is a \emph{real parameter} in the range $ 0\le \epsilon \le 1 $ and $\alpha,\beta$ are \emph{phases}. Importantly, in the universal approach one makes \emph{no assumptions} about how
$\epsilon$, $\alpha$, and $\beta$ depend on the physical control parameters
(e.g., pulse area, detuning, ac Stark shifts, chirp, pulse shape); the only
assumptions are coherent evolution, identical constituent pulses, and accurate
control of their \emph{relative} phases~\cite{GenovUCP,GenovUCP2}. 

A constant phase shift $\varphi$ of the driving field,
$\Omega(t)\rightarrow \Omega(t)e^{i\varphi}$, changes Eq.~(\ref{Uab}) only by
\begin{equation}
\beta \rightarrow \beta+\varphi,
\qquad
\mathbf{U}(\epsilon, \alpha,\beta)\rightarrow \mathbf{U}(\epsilon, \alpha,\beta+\varphi),
\label{phase-shift}
\end{equation}
while $\epsilon$ and $\alpha$ remain unchanged. Therefore, a composite sequence of
$N$ identical pulses with controlled phases $\{\varphi_k\}$ has the propagator
\begin{equation}
\mathbf{U}^{(N)}=
\mathbf{U}\big(\epsilon,\alpha,\beta+\varphi_{N}\big)\cdots
\mathbf{U}\big(\epsilon,\alpha,\beta+\varphi_{2}\big)\,
\mathbf{U}\big(\epsilon,\alpha,\beta+\varphi_{1}\big).
\label{Ucomp}
\end{equation}
By appropriate choice of the phases $\{\varphi_k\}$, one can engineer $\mathbf{U}^{(N)}$
to systematically suppress leading-order sensitivity to imperfections. This is the essence
of the universal composite pulses~\cite{GenovUCP,GenovUCP2}.

In the following sections, we apply this universal parametrization to construct
composite \emph{phase gates} whose performance is intrinsically robust to arbitrary
variations of the driving field parameters while preserving a tunable target phase.


\section{Universal composite phase gates}
\label{UCPG}

The target operation is the phase gate
\begin{equation}
\mathbf{G}(\Phi)= 
\begin{bmatrix}
e^{i\Phi/2} & 0 \\[4pt] 
0 & e^{-i\Phi/2}%
\end{bmatrix}%
,  \label{phase gate}
\end{equation}
which imparts a tunable relative phase $\Phi$ between the qubit amplitudes
$c_{1}$ and $c_{2}$ of Eq.~(\ref{wave function}) while leaving the populations
unchanged. To construct this gate using the universal composite-pulse (UCP)
approach~\cite{GenovUCP,GenovUCP2}, we rely on the propagator
parametrization~(\ref{Uab}), where the real parameter $\epsilon$ quantifies the
deviation from an ideal $\pi$ pulse ($\epsilon=0 \Leftrightarrow |U_{12}|=1$).
In contrast to universal composite $\pi$ pulses, which employ an \emph{odd}
number of pulses to realize robust population inversion, a phase gate must
return the populations to their initial values and therefore requires an
\emph{even} number of nominal $\pi$ pulses. In the ideal limit $\epsilon=0$
each pulse swaps the populations, and an even number of swaps produces the
diagonal propagator~(\ref{phase gate}).

For a pulse with an additional controllable phase $\varphi$, Eq.~(\ref{phase-shift})
shows that only $\beta$ is shifted, $\beta\rightarrow\beta+\varphi$. Hence, without
loss of generality we absorb the (unknown but common) phase $\beta$ into the pulse
phases and write the single-pulse propagator as
\begin{equation}
\mathbf{U}(\epsilon,\alpha;\varphi)=
\begin{bmatrix}
\epsilon\,e^{i\alpha} & \sqrt{1-\epsilon^{2}}\,e^{i\varphi} \\[4pt]
-\sqrt{1-\epsilon^{2}}\,e^{-i\varphi} & \epsilon\,e^{-i\alpha}
\end{bmatrix}.
\label{Uphi}
\end{equation}
A composite sequence of $N$ even identical pulses with phases $\{\varphi_k\}$ then has the propagator 
\begin{equation}
\mathbf{U}^{(N)}=\mathbf{U}(\epsilon,\alpha;\varphi_N)\cdots
\mathbf{U}(\epsilon,\alpha;\varphi_2)\mathbf{U}(\epsilon,\alpha;\varphi_1).
\label{Un}
\end{equation}

\subsection{Four-pulse universal composite phase gate}

The simplest nontrivial universal composite phase gate is obtained with $N=4$ pulses,
\begin{equation}
(\varphi_1,\varphi_2,\varphi_3,\varphi_4)=(0,\phi_1,\phi_2,\phi_3),
\label{phases4}
\end{equation}
where we have set the first phase to zero because only \emph{relative} phases affect the
composite evolution; an overall common phase shift $\varphi_k\rightarrow\varphi_k+\varphi_0$
for all $k$ produces only a global (physically irrelevant) phase factor. This convention is
standard in composite-pulse design and is also used in the universal composite-pulse
construction of Genov \emph{et al.}~\cite{GenovUCP,GenovUCP2}. The parameter $\phi_1$ will
remain free and will be used to control the target gate phase.
In the ideal limit $\epsilon=0$ the propagator of a single pulse becomes
\begin{equation}
\mathbf{U}_0(\varphi)=
\begin{bmatrix}
0 & e^{i\varphi}\\
-e^{-i\varphi} & 0
\end{bmatrix},
\end{equation}
and the four-pulse product is diagonal,
\begin{equation}
\mathbf{U}^{(4)}(\epsilon=0)=
\begin{bmatrix}
e^{i\Lambda} & 0\\
0 & e^{-i\Lambda}
\end{bmatrix},
\qquad
\Lambda=\phi_1-\phi_2+\phi_3.
\label{U0_4}
\end{equation}
Therefore, the ideal action is a phase gate with phase
\begin{equation}
\Phi = 2\Lambda = 2(\phi_1-\phi_2+\phi_3)\quad (\mathrm{mod}\ 2\pi),
\label{PhiLambda}
\end{equation}
which can be tuned by the phases $\phi_k$.

To impose universality, we expand the undesired off-diagonal element
$U^{(4)}_{12}$ in a power series in $\epsilon$ about $\epsilon=0$,
\begin{equation}
U^{(4)}_{12} = \epsilon\,C_1(\alpha;\phi_1,\phi_2,\phi_3) +
\epsilon^3\,C_3(\alpha;\phi_1,\phi_2,\phi_3) + O(\epsilon^5),
\label{U12series}
\end{equation}
where only odd powers appear for an even number of nominal $\pi$ pulses.
A direct multiplication of (\ref{Uphi}) yields for the linear term
\begin{widetext}
\begin{align}
C_1(\alpha;\phi_1,\phi_2,\phi_3)
&= -e^{-2i(\alpha+\phi_1+\phi_2)}
\Big[
e^{i\alpha}\Big(e^{i(\phi_1+2\phi_2+\phi_3)}+e^{i(3\phi_1+\phi_2+\phi_3)}\Big)
+e^{3i\alpha}\Big(e^{i(\phi_1+3\phi_2)}+e^{i(2\phi_1+\phi_2+\phi_3)}\Big)
\Big].
\label{C1_4}
\end{align}
\end{widetext}
Because the phase $\alpha$ is unknown and may depend on \emph{any} driving-field
imperfection~\cite{GenovUCP,GenovUCP2}, universality requires that the coefficients of the
independent harmonics $e^{i\alpha}$ and $e^{3i\alpha}$ vanish separately, i.e.,
\begin{align}
e^{i(\phi_1+2\phi_2+\phi_3)}+e^{i(3\phi_1+\phi_2+\phi_3)} &= 0,
\label{cond1}\\
e^{i(\phi_1+3\phi_2)}+e^{i(2\phi_1+\phi_2+\phi_3)} &= 0.
\label{cond3}
\end{align}
Equations (\ref{cond1})--(\ref{cond3}) admit the compact solution family
\begin{equation}
\phi_2 = 2\phi_1+\pi,\qquad
\phi_3 = 3\phi_1+\pi
\quad (\mathrm{mod}\ 2\pi),
\label{sol4}
\end{equation}
which cancels the entire $O(\epsilon)$ contribution to $U^{(4)}_{12}$ for any $\alpha$.
With (\ref{sol4}), the ideal phase (\ref{PhiLambda}) becomes
\begin{equation}
\Lambda=\phi_1-(2\phi_1+\pi)+(3\phi_1+\pi)=2\phi_1.
\label{Phi4}
\end{equation}
Thus the target phase is tuned simply by choosing
\begin{equation}
\phi_1=\Phi/4 \quad (\mathrm{mod}\ 2\pi).
\label{phi1choice}
\end{equation}
and the corresponding four-pulse universal composite phase gate is
\begin{equation}
(\varphi_1,\varphi_2,\varphi_3,\varphi_4)=
\Big(0,\ \Phi/4,\ \Phi/2+\pi,\ 3\Phi/4+\pi\Big)
.
\label{UCPG4}
\end{equation}

Finally, substituting (\ref{sol4}) into (\ref{U12series}) gives
\begin{equation}
U^{(4)}_{12} =
\epsilon^3\,e^{-3i\alpha}\Big(e^{2i\alpha}-e^{i\phi_1}\Big)\Big(e^{2i\alpha}+e^{i\phi_1}\Big)^2
+O(\epsilon^5),
\label{U12_leading_4}
\end{equation}
showing that the leading leakage amplitude scales as $|U^{(4)}_{12}|\propto \epsilon^3$, uniformly with
respect to arbitrary variations of the physical pulse parameters that are encoded in
$\epsilon$ and $\alpha$. 

\subsection{$N$-pulse universal composite phase gate}
\label{subsec:UCPG_N}

We now present a compact analytic family of \emph{universal composite phase gates} for an
arbitrary \emph{even} number of nominal $\pi$ pulses, $N=2n$. As before, we fix the irrelevant
global phase by setting $\varphi_1=0$, since only relative phases affect the composite
propagator, and we write
\begin{equation}
(\varphi_1,\varphi_2,\ldots,\varphi_N)=(0,\phi_1,\phi_2,\ldots,\phi_{N-1}).
\end{equation}
By solving the universality conditions analytically for $N=4,8,$ and $12$ pulses---via explicit
series expansions of $U^{(N)}_{12}$ about $\epsilon=0$ and cancellation of the lowest-order
coefficients for arbitrary dynamical phase $\alpha$---we found that the resulting solutions are
captured by a simple closed-form phase law. Having identified this pattern, we further verified
numerically that it satisfies the universality conditions for \emph{every} even $N$ up to $N=26$.
In Sec.~\ref{sec:numerics} we illustrate several representative cases, confirming that the same
formula consistently produces broad high-fidelity plateaus and the expected increase in
robustness with increasing~$N$.
The corresponding one-parameter family can be written as

\begin{equation}
\varphi_k=(k-1)\phi_1+\frac{\pi}{n}(k-1)(k-2)\pmod{2\pi},
\label{eq:phase_law_general}
\end{equation}
\begin{equation}
k=1,2,\ldots,N,\qquad N=2n.
\end{equation}

where $\phi_1$ remains a free parameter that sets the target gate phase.

\paragraph*{Ideal phase.}
In the ideal limit $\epsilon=0$ the composite propagator is diagonal,
\begin{equation}
\mathbf{U}^{(N)}(\epsilon=0)=
\begin{bmatrix}
e^{i\Lambda_N} & 0\\
0 & e^{-i\Lambda_N}
\end{bmatrix},
\end{equation}
with
\begin{equation}
\Lambda_N=\varphi_2-\varphi_3+\varphi_4-\cdots+\varphi_N.
\label{eq:LambdaN_alt}
\end{equation}
Substituting Eq.~(\ref{eq:phase_law_general}) into Eq.~(\ref{eq:LambdaN_alt}) yields
\begin{equation}
\Lambda_N=n\phi_1,
\end{equation}
so that the ideal operation is the phase gate $\mathbf{G}(\Phi)$ with
\begin{equation}
\Phi = 2\Lambda_N = N\phi_1 \quad (\mathrm{mod}\ 2\pi).
\label{eq:Phi_general}
\end{equation}
Thus, for a desired target phase $\Phi$ one simply chooses
\begin{equation}
\phi_1=\Phi/N \quad (\mathrm{mod}\ 2\pi),
\end{equation}
and Eq.~(\ref{eq:phase_law_general}) provides the complete set of phases directly in terms
of gate phase $\Phi$:
\begin{equation}
\varphi_k=(k-1)\frac{\Phi}{N}+\frac{2\pi}{N}(k-1)(k-2)\pmod{2\pi},
\label{eq:phase_law_general_Phi}
\end{equation}
\begin{equation}
k=1,2,\ldots,N,\qquad N=2n.
\end{equation}

\paragraph*{Universality order.}
For an even number of nominal $\pi$ pulses the undesired off-diagonal element admits an
odd-power expansion about $\epsilon=0$,
\begin{equation}
U^{(N)}_{12}=\epsilon\,C_1^{(N)}(\alpha)+\epsilon^3\,C_3^{(N)}(\alpha)+\epsilon^5\,C_5^{(N)}(\alpha)+\cdots,
\label{eq:U12seriesN}
\end{equation}
where the coefficients $C_{2j-1}^{(N)}$ are trigonometric polynomials in $e^{i\alpha}$.
The phase law (\ref{eq:phase_law_general}) ensures \emph{universal} cancellation of the
lowest orders---independently of the (unknown) dynamical phase $\alpha$---such that
\begin{equation}
C_{1}^{(N)}=C_{3}^{(N)}=\cdots=C_{2m-1}^{(N)}=0,
\qquad
m=\left\lfloor\frac{N}{4}\right\rfloor,
\label{eq:cancel_orders}
\end{equation}
and therefore
\begin{equation}
U^{(N)}_{12}=O\!\left(\epsilon^{\,2m+1}\right),
\qquad
m=\left\lfloor\frac{N}{4}\right\rfloor.
\label{eq:leading_order_general}
\end{equation}
In particular, Eq.~(\ref{eq:leading_order_general}) reproduces the four-pulse result
$U^{(4)}_{12}=O(\epsilon^3)$ and accurately predicts improved scaling as $N$ increases (e.g.,
$U^{(8)}_{12}=O(\epsilon^5)$ and $U^{(12)}_{12}=O(\epsilon^7)$), while the target phase
remains continuously tunable through $\Phi$ via Eq.~(\ref{eq:Phi_general}).


\section{Numerical simulations}
\label{sec:numerics}

To complement the analytic universality conditions derived in Sec.~\ref{UCPG}, we benchmark
the resulting universal composite phase gates in a concrete physical model and compare them
with the composite phase-gate construction of Torosov \emph{et al.}~\cite{Torosov}.
Specifically, we compare (i) our four-pulse universal composite phase gate with the six-pulse
gate UPh6a, and (ii) our higher-order universal gate (here illustrated with $N=8$) with the
families UPh10a, UPh14a, and UPh26a, where the number indicates the total number of pulses in
the phase-gate sequence. In all cases, the constituent pulses are nominal $\pi$ pulses (equal
nominal pulse areas), and the different robustness properties arise solely from different
choices of the pulse phases.

For clarity, the reference sequences UPh6a, UPh10a, UPh14a, and UPh26a are taken from
Ref.~\cite{Torosov} and correspond to \emph{phase gates constructed from two composite $\pi$
pulses}, i.e., a ``left'' composite $\pi$ sequence followed by an \emph{identical} composite
$\pi$ sequence whose phases are shifted by a constant offset that sets the desired gate phase.
In particular, UPh6a is obtained as a $3+3$ construction (two three-pulse universal $\pi$
sequences U3~\cite{GenovUCP2}), while UPh10a, UPh14a, and UPh26a are obtained analogously as
$5+5$, $7+7$, and $13+13$ constructions, respectively, from universal composite $\pi$ sequences
with 5, 7, and 13 pulses ~\cite{GenovUCP2}. In our numerical benchmarks all sequences---ours and the reference
ones---use the same single-pulse Hamiltonian~(\ref{eq:H_numeric}), the same nominal pulse area
$A=\pi$ for each constituent pulse, and the same target phase $\Phi$ in the fidelity measure
(\ref{eq:F_numeric}); hence the comparison isolates the effect of the phase patterns alone.

We emphasize that the reference UPh6a, UPh10a, UPh14a, and UPh26a sequences are themselves
\emph{universal} composite phase gates within the Torosov construction; thus our comparison is
between \emph{universal} designs with the same nominal pulse area and target phase, and the
observed advantage stems from the different phase patterns, which in our case achieve the same
(or higher) error suppression with fewer pulses and a wider high-fidelity plateau.

\subsection{Model and fidelity measure}

We simulate a driven two-level system within the rotating-wave approximation. During a pulse
with phase $\varphi$ we assume constant Rabi frequency and detuning (rectangular-pulse model),
\begin{equation}
\Omega(t)=\Omega,\qquad \Delta(t)=\Delta\qquad (t_i\le t\le t_f),
\end{equation}
and include the laser phase in the off-diagonal couplings. The Hamiltonian for a pulse reads
\begin{equation}
\mathbf{H}=\frac{\hbar}{2}
\begin{bmatrix}
-\Delta & \Omega e^{i\varphi}\\
\Omega e^{-i\varphi} & \Delta
\end{bmatrix}
=
\frac{\hbar}{2}\Big[\Omega(\cos\varphi\,\sigma_x+\sin\varphi\,\sigma_y)+\Delta\,\sigma_z\Big].
\label{eq:H_numeric}
\end{equation}
Each pulse is characterized by the nominal pulse area $A=\Omega_0T$ and a phase $\varphi$,
where $T$ is the pulse duration and $\Omega_0$ is the nominal (design) Rabi frequency. In the
simulations we set $A=\pi$ for all pulses. Pulse-area (amplitude) errors are introduced by
scaling the Rabi frequency as
\begin{equation}
\Omega=\Omega_0(1+\epsilon_A),
\label{eq:amp_error}
\end{equation}
where $\epsilon_A$ is the relative amplitude error. The detuning is expressed in units of the
nominal Rabi frequency, $\delta=\Delta/\Omega_0$. Thus, the fidelity landscapes are plotted in
the parameter plane $(\epsilon_A,\delta)$.

For constant $\Omega$ and $\Delta$ the single-pulse propagator has a closed form. Defining the
generalized Rabi frequency
\begin{equation}
\Omega_{\mathrm{eff}}=\sqrt{\Omega^2+\Delta^2},
\end{equation}
the rotation angle accumulated during the pulse is
\begin{equation}
\theta=\Omega_{\mathrm{eff}}T
=
A\sqrt{(1+\epsilon_A)^2+\delta^2},
\label{eq:theta_numeric}
\end{equation}
and the rotation axis is
\begin{equation}
\hat{\mathbf{n}}=\frac{1}{\sqrt{(1+\epsilon_A)^2+\delta^2}}
\Big((1+\epsilon_A)\cos\varphi,\ (1+\epsilon_A)\sin\varphi,\ \delta\Big).
\label{eq:n_numeric}
\end{equation}
The corresponding unitary for a single pulse is
\begin{equation}
\mathbf{U}(A,\varphi;\epsilon_A,\delta)
=
\cos\!\Big(\frac{\theta}{2}\Big)\mathbf{1}
-i\sin\!\Big(\frac{\theta}{2}\Big)\big(\hat{\mathbf{n}}\cdot\bm{\sigma}\big).
\label{eq:U_numeric}
\end{equation}

A composite phase gate is specified by the phase list $\{\varphi_k\}_{k=1}^{N}$ (and here a
common nominal area $A=\pi$). The total propagator is obtained by ordered multiplication,
\begin{equation}
\mathbf{U}_{\mathrm{tot}}=
\mathbf{U}(A,\varphi_N)\cdots \mathbf{U}(A,\varphi_2)\mathbf{U}(A,\varphi_1),
\label{eq:Utot_numeric}
\end{equation}
where all pulses share the same $(\epsilon_A,\delta)$.

To quantify the performance we use a global-phase-insensitive fidelity based on the unitary
overlap between the actual propagator $\mathbf{U}_{\mathrm{tot}}$ and the ideal target phase gate
$\mathbf{G}(\Phi)$. Since a global phase has no physical effect, we define
\begin{equation}
F=\frac{1}{2}\left|\mathrm{Tr}\!\left[\mathbf{G}^{\dagger}(\Phi)\,\mathbf{U}_{\mathrm{tot}}\right]\right|,
\label{eq:F_numeric}
\end{equation}
which satisfies $F=1$ if and only if $\mathbf{U}_{\mathrm{tot}}=e^{i\chi}\mathbf{G}(\Phi)$ for some
global phase $\chi$. In the figures we plot contours of the infidelity $1-F(\epsilon_A,\delta)$,
which delineate the high-fidelity operating regions. 

We note that the resulting fidelity landscapes are essentially \emph{independent} of the target
phase~$\Phi$: changing~$\Phi$ mainly shifts the overall acquired phase while leaving the shape and
extent of the high-fidelity plateau in the $(\epsilon_A,\delta)$ plane practically unchanged.
This phase-independence is a direct consequence of the universal construction, which enforces
cancellation of the leading error terms for arbitrary dynamical phase~$\alpha$, and therefore does
not rely on any particular choice of the target gate phase.

\subsection{Parameter scan and comparison}

For each gate design we evaluate $F$ on a uniform $100\times100$ grid in the parameter plane,
\begin{equation}
\epsilon_A\in[-0.3,0.3], \qquad \delta\in[-0.5,0.5],
\end{equation}
i.e.\ $10^4$ parameter points per sequence. At each grid point we compute the single-pulse
propagators via Eq.~(\ref{eq:U_numeric}), form $\mathbf{U}_{\mathrm{tot}}$ by Eq.~(\ref{eq:Utot_numeric}),
and evaluate the fidelity by Eq.~(\ref{eq:F_numeric}).

\begin{figure}
\centering
\includegraphics[width=\columnwidth]{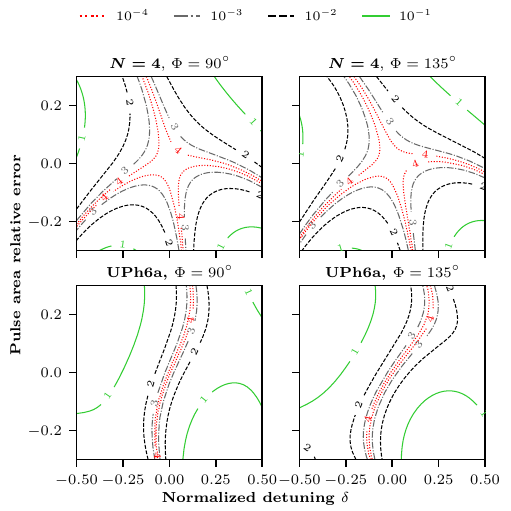}
\caption{Numerical fidelity landscapes for the target phase gate $\mathbf{G}(\Phi)$ for our
four-pulse universal composite phase gate (top) and the universal six-pulse gate UPh6a of
Torosov \emph{et al.}~\cite{Torosov} (bottom). The contours labeled $m$ indicate fixed
infidelity $1-F=10^{-m}$.}
\label{fig:num_4_vs_33}
\end{figure}

Figure~\ref{fig:num_4_vs_33} compares our universal four-pulse sequence with the universal six-pulse
sequence UPh6a from Ref.~\cite{Torosov} for two target phases, $\Phi=90^{\circ}$ and
$\Phi=135^{\circ}$. The reference UPh6a can be viewed as a concatenation of two universal
$\pi$-pulse sequences (U3)~\cite{GenovUCP2} with three pulses each. In agreement with our
construction, the robustness profile of the four-pulse gate is essentially independent of
$\Phi$, because the analytic conditions enforce cancellation for arbitrary dynamical phase
$\alpha$. The resulting high-fidelity plateau is substantially broader than that of UPh6a,
consistent with the leading leakage scaling $U^{(4)}_{12}=O(\epsilon^3)$.

\begin{figure}
\centering
\includegraphics[width=\columnwidth]{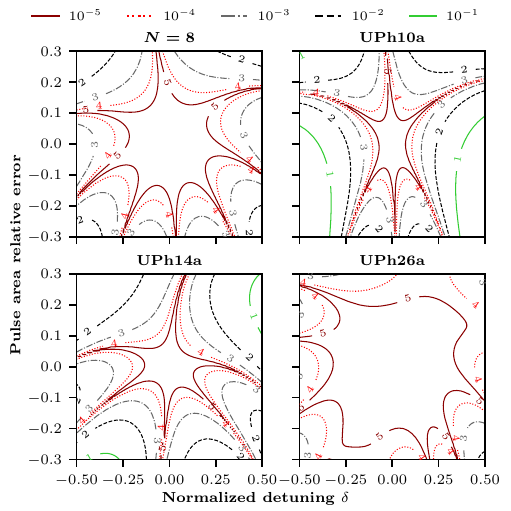}
\caption{Numerical fidelity landscapes for the target phase gate $\mathbf{G}(\Phi)$ for our
eight-pulse universal composite phase gate (top left) compared with phase-gate constructions
based on pairs of universal $\pi$-pulse sequences (UPh10a, UPh14a, UPh26a)~\cite{Torosov}.
Contour labels indicate fixed infidelity $1-F=10^{-m}$.}
\label{fig:num_10_pulse}
\end{figure}

Figure~\ref{fig:num_10_pulse} compares our $N=8$ universal composite phase gate with the universal
phase-gate sequences UPh10a, UPh14a, and UPh26a~\cite{Torosov} for a representative target phase
$\Phi=90^{\circ}$. Although our construction uses fewer constituent pulses than UPh14a, it
produces a visibly broader high-fidelity plateau under \emph{simultaneous} pulse-area and
detuning errors. This enhanced robustness is consistent with the higher-order suppression of
leakage predicted by the universal expansion, namely $U^{(8)}_{12}=O(\epsilon^{5})$.

\begin{figure}
\centering
\includegraphics[width=\columnwidth]{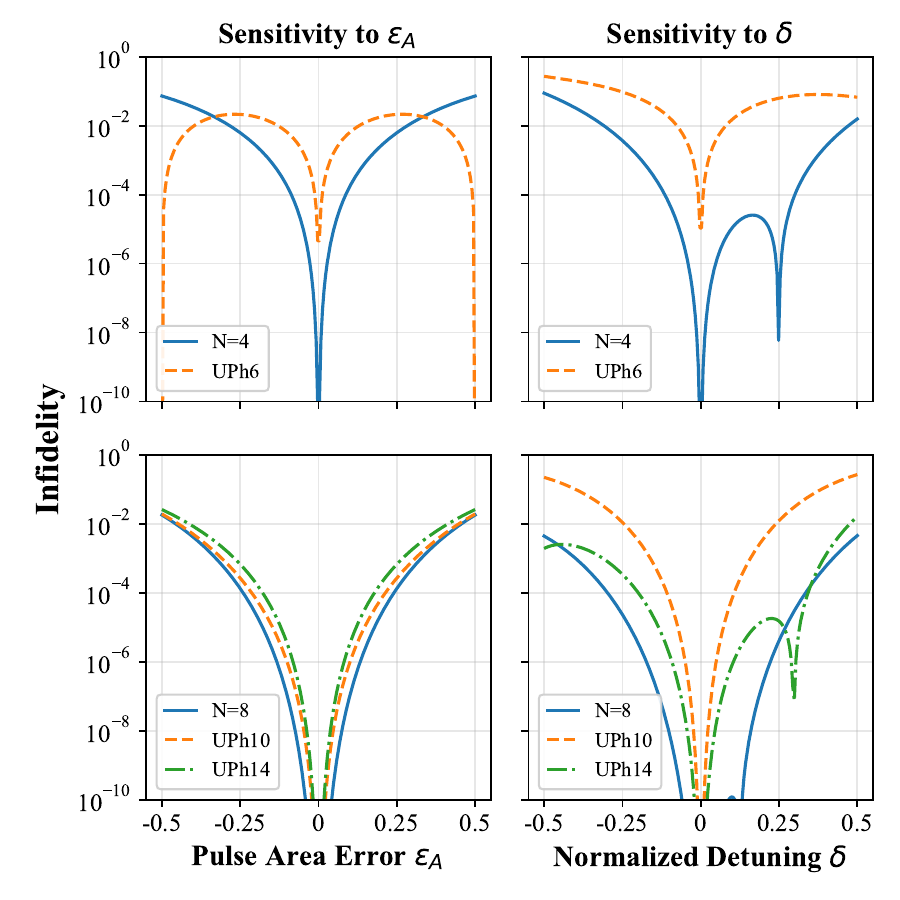}
\caption{One-dimensional cross-sections of the infidelity for our universal composite phase gates
compared with the Torosov sequences. Top row: $N=4$ (solid blue) versus UPh6a (dashed orange).
Bottom row: $N=8$ (solid blue) versus UPh10a (dashed orange) and UPh14a (dash-dotted green).
Left column: infidelity versus pulse-area error $\epsilon_A$ at $\delta=0$.
Right column: infidelity versus normalized detuning $\delta$ at $\epsilon_A=0$.}
\label{fig:1d_cuts}
\end{figure}

To provide a more quantitative comparison, Fig.~\ref{fig:1d_cuts} shows one-dimensional
cross-sections of the infidelity along the $\epsilon_A=0$ and $\delta=0$ axes. The top row
confirms that our four-pulse gate achieves a broader high-fidelity region than the six-pulse
UPh6a, while the bottom row demonstrates that our eight-pulse gate outperforms UPh10a and
rivals UPh14a with only 8 pulses versus 14.

\begin{figure}
\centering
\includegraphics[width=\columnwidth]{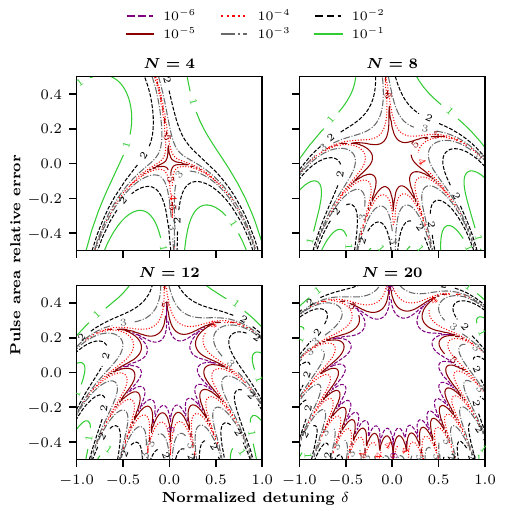}
\caption{Numerical fidelity landscapes for the target phase gate $\mathbf{G}(\Phi)$ for our
universal composite phase gates of 4, 8, 12 and 20 pulses constructed using the general phase formula \eqref{eq:phase_law_general_Phi}.
Contour labels indicate fixed infidelity $1-F=10^{-m}$.}
\label{fig:4_8_12_and_20_pulse}
\end{figure}

To validate the performance of the general phase prescription \eqref{eq:phase_law_general_Phi},
Fig.~\ref{fig:4_8_12_and_20_pulse} shows infidelity contour maps for the $N=4,8,12,$ and $20$
universal composite phase gates at a representative target phase $\Phi=\pi/4$.
In accordance with Eq.~(\ref{eq:leading_order_general}), these sequences cancel the leakage
terms through $O(\epsilon)$, $O(\epsilon^{3})$, $O(\epsilon^{5})$, and $O(\epsilon^{9})$,
respectively, so that the leading residual scales as
$U^{(4)}_{12}=O(\epsilon^{3})$, $U^{(8)}_{12}=O(\epsilon^{5})$,
$U^{(12)}_{12}=O(\epsilon^{7})$, and $U^{(20)}_{12}=O(\epsilon^{11})$.

Finally, we emphasize that the rectangular-pulse assumption (constant $\Omega$ and $\Delta$
during each pulse) is adopted solely to provide a transparent and reproducible numerical
benchmark. The universal composite phase gates derived in Sec.~\ref{UCPG} are formulated at the
level of the single-pulse propagator and do not rely on any specific pulse envelope;
therefore, using smooth pulse shapes leads to qualitatively similar fidelity landscapes, with
the same characteristic high-fidelity plateaus and robustness trends.

\subsection{Validation with shaped pulses}

In Sec.~\ref{UCPG} we derived the composite phase sequences at the level of the single-pulse propagator and
therefore claimed universality with respect to experimental imperfections \emph{independently} of the detailed
pulse envelope. The numerical results above were obtained in the rectangular-pulse model and by
evaluating the propagator analytically in closed form. To explicitly validate (i) that the robustness persists
for a smooth pulse shape and (ii) that the expected behavior is recovered when the dynamics are simulated
directly, we repeat the parameter scans using a time-domain solution of the Schr\"odinger equation for each
constituent pulse rather than Eq.~(\ref{eq:U_numeric}). Because the shaped-pulse Hamiltonian is not identical
to the rectangular benchmark, the high-fidelity plateaus acquire a somewhat different shape; nevertheless,
they exhibit the same qualitative robustness trends and extended high-fidelity regions, confirming that the
observed performance is a property of the composite phase patterns rather than an artifact of the closed-form
rectangular propagator.

\begin{figure}
\centering
\includegraphics[width=\columnwidth]{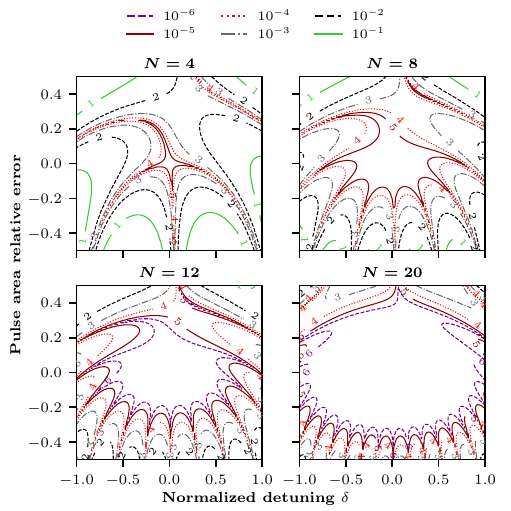}
\caption{Numerical fidelity landscapes for the target phase gate \(\mathbf{G}(\Phi)\) for our universal composite
phase gates (same phase patterns as Fig.~\ref{fig:4_8_12_and_20_pulse}), computed by solving the time-dependent
Schr\"odinger equation with truncated-Gaussian pulse envelopes. Contour labels indicate fixed infidelity
\(1-F=10^{-m}\).}
\label{fig:num_gauss_sim}
\end{figure}

Concretely, we replace the constant-\(\Omega\) assumption by a truncated Gaussian Rabi envelope while keeping
the same nominal pulse area \(A=\int_{t_i}^{t_f}\Omega(t)\,dt=\pi\) for every pulse. For a pulse of duration
\(T\) we take \(\Omega(t)=\Omega_0(1+\epsilon_A)\,g(t)\), where \(g(t)\) is a Gaussian centered at \(T/2\) with
standard deviation \(\sigma=0.18\,T\), and we truncate the tails by setting \(g(t)=0\) for
\(|t-T/2|>2.5\,\sigma\). The envelope is normalized so that its time average over \([0,T]\) equals unity,
ensuring that the nominal pulse area remains \(A=\Omega_0T=\pi\) and thus making the comparison directly
analogous to the rectangular case. The composite propagator is then obtained by time-ordering the evolution
under the Hamiltonian~(\ref{eq:H_numeric}) with this shaped \(\Omega(t)\), and the fidelity is evaluated using
Eq.~(\ref{eq:F_numeric}). The corresponding contour maps are shown in Fig.~\ref{fig:num_gauss_sim}.

\section{Conclusions}

We have developed a systematic construction of \emph{universal} composite
phase gates based on the general Cayley--Klein parametrization of the single-pulse
propagator. In contrast to universal composite $\pi$ pulses, which employ an odd number
of pulses to achieve robust population inversion, our phase-gate designs use an \emph{even}
number of nominal $\pi$ pulses so that the populations return to their initial values
while a prescribed relative phase is accumulated. Universality is obtained by expanding
the undesired off-diagonal element $U_{12}$ in a power series about $\epsilon=0$ and
imposing order-by-order cancellation of the leading coefficients for arbitrary dynamical
phase $\alpha$, without making any assumptions about how the physical imperfections map
onto $(\epsilon,\alpha)$. As a result, the constructed gates are intrinsically robust to
variations in any driving-field parameter (pulse area, detuning, ac Stark shifts, chirp,
etc.) and are applicable with arbitrary pulse shapes.

A central outcome of this work is a compact analytic phase prescription for an arbitrary even number of pulses $N$, which produces a one-parameter family of \emph{universal} composite phase gates with a continuously tunable target phase $\Phi$ [Eqs.~(\ref{eq:phase_law_general}) and (\ref{eq:phase_law_general_Phi})]. Motivated by the explicit closed-form solutions obtained for $N=4,8,$ and $12$ pulses, we infer a general expression that generates the phase sets for any even $N$. We have further validated this prescription by explicit evaluation of the composite propagator for all even pulse numbers up to $N=26$. The resulting sequences suppress the dominant leakage contribution to $U_{12}$ to high order, yielding broad, flat high-fidelity plateaus while preserving full tunability of the phase-gate angle.

Finally, our numerical simulations in Sec.~\ref{sec:numerics} confirm the predicted robustness
and demonstrate a clear performance advantage over the composite phase-gate constructions of
Torosov \emph{et al.}~\cite{Torosov}. For the same target phase, our four-pulse gate exhibits a
substantially wider high-fidelity region than the universal six-pulse UPh6a sequence, and our higher-order
designs achieve robustness comparable to, or better than, much longer Torosov-type sequences.
These results establish our universal composite phase gates as a simple and versatile route to
high-fidelity, error-resilient phase control in quantum information processing and related
coherent-control applications.


\acknowledgments
This research is partially supported by the Bulgarian national plan for
recovery and resilience, contract BG-RRP-2.004-0008-C01 SUMMIT: Sofia
University Marking Momentum for Innovation and Technological Transfer,
project number 3.1.4.



\begin{thebibliography}{99}
\bibitem{QI} M. A. Nielsen and I. L. Chuang, \emph{Quantum Computation and
Quantum Information} (Cambridge University Press, 1990).

\bibitem{Shor1} P. W. Shor, in Proceedings of the 35th Annual Symposium on
the Foundations of Computer Science, edited by S. Goldwasser (IEEE Computer
Society, Los Alamitos, 1994), p. 124.

\bibitem{Shor2} P. W. Shor, SIAM J. Sci. Stat. Comput. \textbf{26}, 1484
(1997).

\bibitem{Grover} L. K. Grover, Phys. Rev. Lett. \textbf{79}, 325 (1997).

\bibitem{Cirac} T. Calarco, D. Jaksch, J. I. Cirac and P. Zoller, J. Opt. B 
\textbf{4}, 430 (2002).

\bibitem{geometric1} M. V. Berry, Proc. R. Soc. London, Ser. A \textbf{392},
45 (1984).

\bibitem{geometric2} R. G. Unanyan, B. W. Shore, and K. Bergmann, Phys. Rev.
A \textbf{59}, 2910 (1999).

\bibitem{geometric3} R. Unanyan, M. Fleischhauer, B. W. Shore, K. Bergmann,
Opt. Commun. \textbf{155}, 144 (1998).

\bibitem{geometric4} A. Ekert, M. Ericsson, P. Hayden, H. Inamori, J. A.
Jones, D. K. L. Oi and V. Vedral, J. Mod. Opt. \textbf{47}, 2501 (2000).

\bibitem{laser phases1} H. Goto and K. Ichimura, Phys. Rev. A \textbf{70},
012305 (2004).

\bibitem{laser phases2} X. Lacour, S. Gu\'erin, N. V. Vitanov, L. P.
Yatsenko, and H. R. Jauslin, Opt. Commun. \textbf{264}, 362 (2006).

\bibitem{Levitt79} M. H. Levitt and R. Freeman, J. Magn. Reson. \textbf{33},
473 (1979).

\bibitem{Freeman80} R. Freeman, S. P. Kempsell, and M. H. Levitt, J. Magn.
Reson. \textbf{38}, 453 (1980).

\bibitem{Levitt86} M. H. Levitt, Prog. Nucl. Magn. Reson. Spectrosc. \textbf{%
18}, 61 (1986).

\bibitem{Freeman97} R. Freeman, \emph{Spin Choreography} (Spektrum, Oxford,
1997).

\bibitem{GenovUCP} G. T. Genov, D. Schraft, T. Halfmann, and N. V. Vitanov,
Phys. Rev. Lett. \textbf{113}, 043001 (2014).

\bibitem{GenovUCP2} G. T. Genov, M. Hain, N. V. Vitanov, and T. Halfmann,
Phys. Rev. A \textbf{101}, 013827 (2020).

\bibitem{Torosov} B. T. Torosov and N. V. Vitanov, Phys. Rev. A \textbf{90},
012341 (2014).

\bibitem{Gevorgyan} H. L. Gevorgyan and N. V. Vitanov Phys. Rev. A \textbf{109},
052625 (2024).


\bibitem{Allen-Eberly} L. Allen and J. H. Eberly, \emph{Optical Resonance
and Two-Level Atoms} (Wiley, New York, 1975).

\bibitem{Shore} B. W. Shore, \emph{The Theory of Coherent Atomic Excitation}
(Wiley, New York, 1990).

\end{thebibliography}
\end{document}